\documentclass{caosp309}

\usepackage{graphicx}

\usepackage{natbib}
\articleNo{301}
\pubyear{2020}
\volume{50}
\volnumber{3}
\firstpage{1}
\received{May 1, 2020}
\accepted{July 28, 2020}

\def\BibTeX{{\rm B\kern-.05em{\sc i\kern-.025em b}\kern-.08em
             T\kern-.1667em\lower.7ex\hbox{E}\kern-.125emX}}

\newcommand{\creme}{CRÉME}

\begin{document}

%
\htitle{Adding TESS to CRÉME}
\hauthor{K.G.\,He{\l}miniak, {\it et al.}}

\title{Adding TESS to CRÉME}
\subtitle{Light curves and masses of 300+ eclipsing binaries}

%
%
\author{
       K.G.\,He{\l}miniak\inst{1}\orcid{0000-0002-7650-3603}
      \and
        A.\,Moharana\inst{1,2}\orcid{0000-0002-9616-512X} 
      \and 
        T.B.\,Pawar\inst{1}\orcid{0000-0002-0004-0569}  
      \and 
        G.\,Pawar\inst{1}\orcid{0000-0003-3639-9052}
       }

%
\institute{
           Nicolaus Copernicus Astronomical Center, Polish Academy of Sciences, ul. Rabia\'nska 8, 87-100 Toru\'n, Poland, \email{xysiek@ncac.torun.pl}
         \and 
           Astrophysics Group, Keele University, Staffordshire, ST5 5BG, UK
          }

\date{November 15, 2024}

\maketitle

\begin{abstract}
The Comprehensive Research with Echelles on the Most interesting Eclipsing binaries (\creme) projects was aimed to collect high-resolutions spectra of about 380 detached eclipsing binaries (DEBs), which mostly do not have literature RV data. From this vast observational material we were able to estimate masses of components of 325 double-lined system. Since the launch of the TESS mission we have been collecting 2-min cadence photometry for the \creme\ targets through successful GI proposals. As by Sector 85, we obtained data for $>$330 of them. We are thus now in the process of comprehensively analysing our targets. This paper presents the recent status of the \creme\ project and its space photometry counterpart, and describes several sub-projects within \creme\ that focus on specific classes of targets.
\end{abstract}

\section{Introduction}\label{intro}
Detached eclipsing binaries (DEBs), especially those that are also double-lined spectroscopic (SB2) pairs, are one of the most useful objects in astrophysics. Their importance can not be overestimated, as they are used, for example to: test models of stellar structure and evolution, derive observational calibrations, test stellar formation theories, provide high-precision distances, characterise extrasolar planet hosts, etc. Many crucial stellar parameters, like mass, radius, effective temperature, age, or metallicity, can be derived simultaneously and accurately only from DEBs. 
So far, fewer than 350 DEBs have the desired level of uncertainty in masses or radii ($<$2\%), and a large fraction still does not have reliable information about the age and metallicity \citep{sout15}. Furthermore, many interesting classes of stars, and many areas of the H-R diagram, are underrepresented among the best-measured stars. For this reason, a large observational survey, dedicated for searching and characterising interesting stars in DEBs, has been started.

\section{The \creme\ status}

\subsection{Spectroscopy}

\begin{table}
\centering
\scriptsize
\caption{Statistics of telescope time granted for the \creme\, project.\label{tab_stat}}
\begin{tabular}{cc|cc|cc}
\hline \hline
Telesc./Spectr.& Time & Telesc./Spectr.& Time &Telesc./Spectr.& Time \\
 \hline
OAO-188cm/HIDES		& 87.5 n	& TNG/HARPS-N		&	16 n	& Magellan-Clay/PSF & 4 n\\
SMARTS 1.5m/CHIRON	&  753 h	& SALT/HRS			&	116 h	& NOT/FIES		&	4 n	 \\
Euler/CORALIE		&	40 n	& ESO 3.6m/HARPS	&	10 n	& OHP 1.9m/SOPHIE & 3 n  \\
MPG-2.2m/FEROS		&	30 n	& Subaru/HDS+IRCS	&	9 n		& VLT/UVES		&	3.5 h\\
\hline
\end{tabular}
\\ Note: With additional data from AAT/UCLES, Radcliffe/GIRAFFE, Keck~I/HIRES, TNG/SARG, Hamilton/HamSpec (all pre-2011), OUC-50cm/PUCHEROS; and archives: ESO, SOPHIE, ELODIE, KOA, APOGEE.
\end{table}

The {\it Comprehensive Research with Échelles on the Most interesting Eclipsing binaries} (\creme) is an observational project of high-resolution spectroscopic observations, intended to derive precise radial velocity (RV) measurements and atmospheric parameters. The main goals of the project are:
\begin{enumerate}
\item Identification of new examples of rare, poorly studied, or otherwise interesting~DEBs.
\item Precise characterisation of the studied systems i.e. determination of masses, radii, temperatures, distances, metallicities, and ages of stars.
\end{enumerate} 

The \creme\ project was a successor of a previous search for low-mass stars (2005-2010), and a small-scale survey of F- and G-type DEBs (2008-2009) made with iodine cell spectrographs. The project goals were defined and the observations started in 2011. Initially, the target sample was made of DEBs from the southern hemisphere, identified by the All-Sky Automated Survey \citep[ASAS;][]{asas}. In 2013 \creme\, has been extended to the northern hemisphere, with a significant portion of targets coming from the main field of the {\it Kepler} mission. The intensive spectroscopic campaign lasted till 2017, but since then additional data have been taken for selected objects, if needed. 

\begin{figure}
\centering
\includegraphics[width=0.58\textwidth]{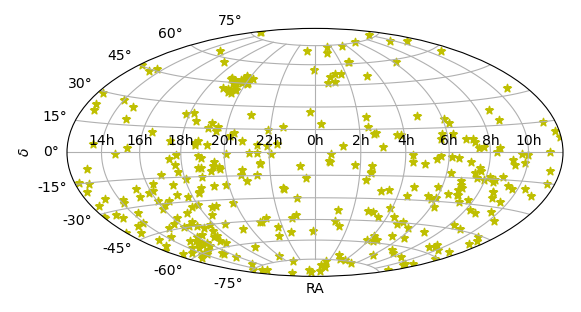}
\includegraphics[width=0.40\textwidth]{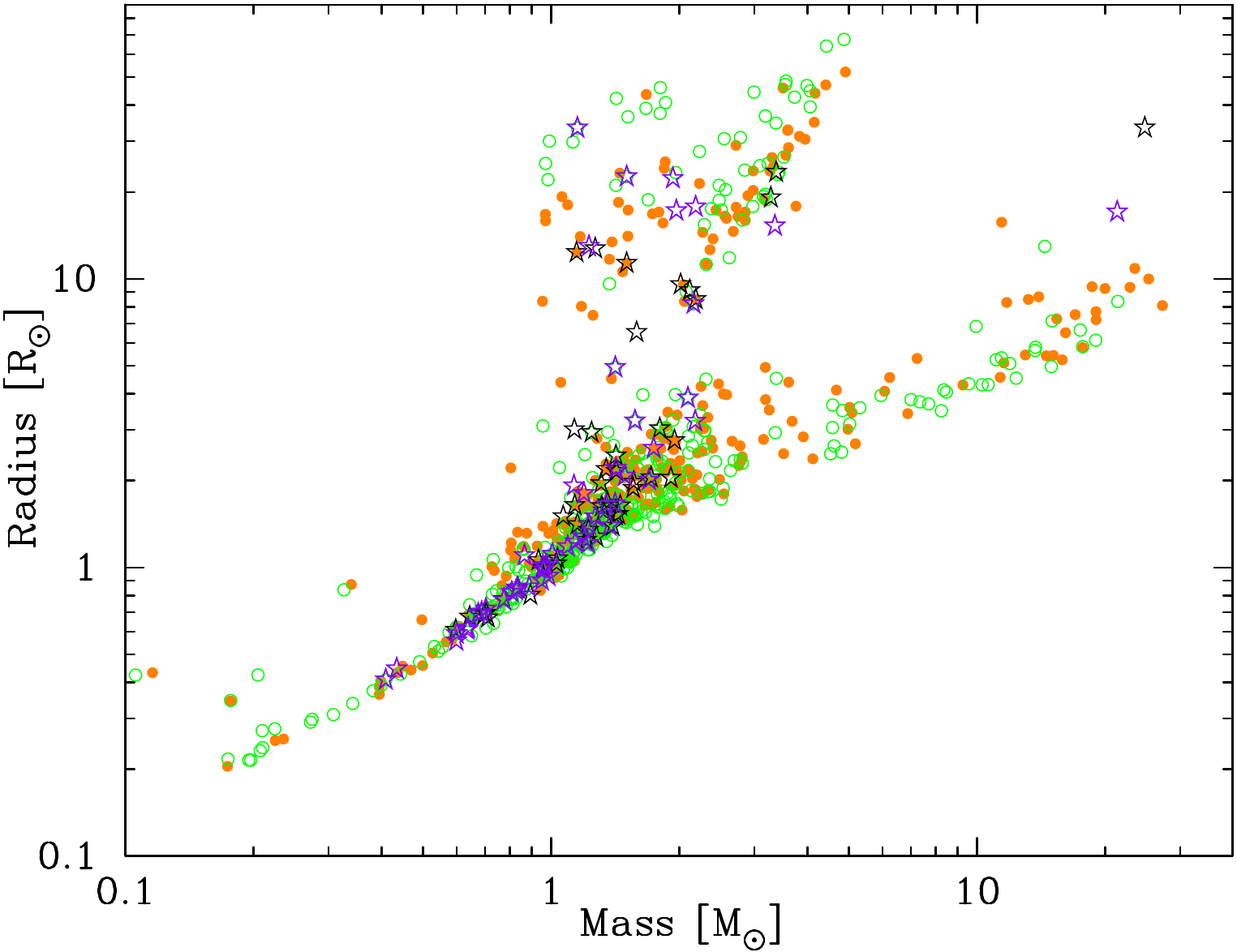}
\caption{Left: On-sky distribution of \creme\ targets.
Right: Masses and radii of binary components from the DEBCat catalogue (orange and green for primaries and secondaries, respectively) and \creme\, DEBs published so far (black and purple for primaries and secondaries, respectively).} 
 \label{fig_cremedebcat}
\end{figure}

At the moment (November 2024) the \creme\, target sample consists of 386 DEBs (Figure~\ref{fig_cremedebcat}). A total of $>$7000 spectra has been gathered with 19 spectrographs attached to 17 telescopes (from 0.5 to 10-m apertures), or were found in public archives. The total telescope time granted to the project exceeded 300 nights (Table~\ref{tab_stat}). 

The project intended to observe relatively bright ($V<12.5$ mag) and ``red'' ($V-K>1.1$ mag) systems, so a sufficient precision of RVs would be possible to obtain. However, several early-type objects with large $E(B-V)$ were included and later kept in the sample. Out of the 386 systems observed within \creme, 32 were marked as not qualified for further studies, 22 were found to be single-lined (SB1), 3 early-type systems pose difficulties in RV determination and require an individual approach, and 4 systems do not have enough sufficient RV measurements ($N<4$). Therefore, we currently have 325 systems with a sufficient number of good-quality data to estimate orbital parameters and masses of the components\footnote{Technically only the lower mass limits $M \sin^3(i)$, but in DEBs $\sin^3(i)\simeq 1$, therefore they are good proxies of true stellar masses.}.

Masses alone, or even the orbital solutions, can already point towards some of the interesting properties, such as: location on the cool ($<$0.9~M$_\odot$) or hot ($>$3~M$_\odot$) parts of the main sequence; location inside the instability strips ($\gamma$~Dor, $\delta$~Sct), or multiplicity (especially in hierarchical triple SB+3, or quadruple SB+SB configurations). In \creme\, we have so far 76 individual low-mass and 19 high-mass stars, 14 confirmed and 8 candidates for pulsating stars in DEBs, or 80 multiples of various architectures, including compact hierarchical triples (outer period $P_3<1000$~d; CHTs) and SB+SB quadruples.

\subsection{TESS, {\it Kepler}, and ground-based photometry}

To obtain the required precision in parameters, especially radii, one needs photometric measurements taken with high enough precision, sufficient cadence, and phase coverage. It quickly became clear that the publicly available data from such projects as ASAS or SuperWASP are not enough for the purposes of \creme, so additional dedicated observations had to be taken. This turned out to be very time-consuming and ineffective, especially when comes to the orbital phase coverage. Only the objects from the {\it Kepler/K2} fields had sufficient data available. Therefore, after the launch TESS, we started applying for 2-minute-cadence photometry of the \creme\, targets through the TESS Guest Investigator (GI) programs. These programs allow to dedicate part of the mission's resources to scientific research outside of the main mission goals. We have been successful during all seven cycles of the mission.

At the moment of writing this article, the TESS satellite is observing in Sector 85 during its seventh cycle of operations. A total of 329 \creme\, targets have their 2-minute cadence photometry, including: 66 in just one sector, 117 in two sectors, 49 in three, 34 in four, 50 in 5 to 12 sectors, and 13 in at least 13 sectors. Additionally, 6 more targets were in the satellite's field of view, thus the longer cadence data (30 or 10 minutes) are available, and another 13 systems are expected to be observed by the end of Cycle 7. Since TESS has by now covered almost the entire sky, including the Ecliptic, we were able to gather data for systems previously observed with {\it Kepler} and K2. Only 3 \creme\ targets have K2 photometry, but no TESS.

\section{\creme\ + TESS (+ {\it Kepler}/K2) results}

In Figure~\ref{fig_cremedebcat} we present masses and radii of components of 348 binaries collected in the DEBCat catalogue, together with 50 systems with published\footnote{Including preliminary solutions presented on conferences, and results based on pre-2011 observations.} orbital and physical parameters based on \creme\, data. One can note on the mass-radius panel that the \creme\, targets appear in regions poorly populated by the DEBCat objects, which shows that the goals of the project are possible to achieve. Notably, about 30 \creme\, targets are now listed in DEBCat, thus some points may overlap. For most of them the solutions utilise TESS photometry from our GI programmes, or {\it Kepler}/K2 data. 
Currently, about 100 DEBs have been modelled with the use of \creme\ data. This includes results presented in 35 publications, as well as unpublished ones (in preparation).

\subsection{Projects dedicated to specific classes of systems}
A number of studies dedicated to specific scientific cases have been conducted by the team. Two of them were described in other contributions in these proceedings: by A.~Moharana \citep[see also:][]{moha23,moha24} and G. Pawar. Other projects include but are not limited to:

\textbf{Masses of 650 stars.}
Using spectroscopic data alone we were able to estimate absolute masses of components of 325 binaries (650 individual stars). From them we selected 78 long-period ($P_{\rm orb}>4$~d) pairs to determine realistic mass precision, unaffected by stellar rotation or activity. We found mass uncertainty lower than 1.0\% in 90\% cases, and for $\sim$60\% of stars was better than 0.5\%. In the best case we reached 0.051+0.058\%.

\textbf{Low-mass stars.}
The \creme\ survey partially evolved from the search for new DEBs composed of low-mass stars \citep[K- and M-type dwarfs, e.g.][]{hel11b}. This topic has been continued, and with the mass estimates of the 650 individual stars we were able to identify a total 19 cases DEBs with both components $M<0.9$~M$_\odot$ (He{\l}miniak et al., in prep.), 29 more cases of pairs with a more massive (F/G-type) primary and low-mass secondary, and 9 more systems that seemingly underwent a mass transfer, where the secondary low-mass component was stripped of its outer layers (Pawar et al., in prep.).

\textbf{Giants and sub-giants.}
This was a topic of high interest in the earlier stages of the project, since before 2010 the DEBCat listed only a few cases. Our publications include both multi-object \citep{rata13,rata16,rata21} and single-object \citep{hel15}. The latter includes HD~187669, the first Galactic double-giant DEB with high-precision stellar parameters. Several othe cases were included in various publications. There are at least 5 more unpublished double-giant systems in \creme\, and numerous others with a sub-giant component.

\textbf{$\delta$~Sct and $\gamma$~Dor pulsators.}
Both kinds of pulsators are main sequence stars of a particular range of masses. By focusing on stars with those mass ranges we selected a number of candidate DEBs and inspected their light curves in search for pulsations. This way we identified and studied 12 systems with $\delta$~Sct-type pulsations \citep[e.g.][]{bpaw24}, some of which are pending publication (in prep.), as well as 6 with $\gamma$~Dor-type components (Pawar et al. in prep).
 
\textbf{Systems with total eclipses.}
The totally-eclipsing DEBs are pairs with a fortunate geometry, when during one of the minima the eclipsed component hides completely behind the other one. For a brief moment we may record light from only one star of the system. This provides several advantages: strong bonds on the flux ratio, individual color indices, or spectroscopic analysis of one of the components without disentangling. These have been utilised in \creme\ studies several times, most recently in \citet{hel24}, where we presented results of a dedicated UVES campaign, aimed for spectra taken during total eclipses of 11 DEBs. 
We showed that by adding spectroscopic information about just one component, we can drastically reduce the uncertainty in age and distance determination.

\acknowledgements
\creme\ is a collaborative project that involved a large number of people, therefore we'd like to thank: R.~Brahm, J.~Coronado, N.~Espinoza, D.~Graczyk, M.~Hempel, A.~Jord\'an, E.~Kambe, M.~Konacki, S.~Koz{\l}owski, H.~Maehara, F.~Marcadon, D.~Minniti, E.~Niemczura, J.~Olszewska, B.~Pilecki, M.~Rabus, M.~Ratajczak, P.~Sybilski, A.~Tajitsu, M.~Tamura, A.~Tokovinin, N.~Ukita, and L.~Vanzi for their various valuable contributions. We'd also like to thank the staff of La Silla-Paranal, CTIO, Okayama, Las Campanas, Subaru, NOT, TNG, OHP, SALT, and AAO telescopes/observatories for their support during many years. 

We acknowledge the support provided by the Polish National Science Center through grants no. 2021/41/N/ST9/02746 and 2023/49/B/ST9/01671. A.M. acknowledges support from the UK Science and Technology Facilities Council (STFC) under grant number ST/Y002563/1. Polish participation in SALT is funded by grant No. MEiN nr 2021/WK/01.

\bibliography{Helminiak_CREME_2024}

\begin{thebibliography}{11}
\expandafter\ifx\csname natexlab\endcsname\relax\def\natexlab#1{#1}\fi

\bibitem[{{He{\l}miniak} {et~al.}(2015){He{\l}miniak}, {Graczyk}, {Konacki},
  {Pilecki}, {Ratajczak}, {Pietrzy{\'n}ski}, {Sybilski}, {Villanova}, {Gieren},
  {Pojma{\'n}ski}, {Konorski}, {Suchomska}, {Reichart}, {Ivarsen}, {Haislip},
  \& {LaCluyze}}]{hel15}
{He{\l}miniak}, K.~G., {Graczyk}, D., {Konacki}, M., {et~al.}, {Orbital and
  physical parameters of eclipsing binaries from the ASAS catalogue - VIII. The
  totally eclipsing double-giant system HD 187669}. 2015, {\it \mnras}, {\bf
  448}, 1945, DOI: 10.1093/mnras/stu2680

\bibitem[{{He{\l}miniak} {et~al.}(2011){He{\l}miniak}, {Konacki},
  {Z{\l}oczewski}, {Ratajczak}, {Reichart}, {Ivarsen}, {Haislip}, {Crain},
  {Foster}, {Nysewander}, \& {Lacluyze}}]{hel11b}
{He{\l}miniak}, K.~G., {Konacki}, M., {Z{\l}oczewski}, K., {et~al.}, {Orbital
  and physical parameters of eclipsing binaries from the All-Sky Automated
  Survey catalogue. III. Two new low-mass systems with rapidly evolving spots}.
  2011, {\it \aap}, {\bf 527}, A14, DOI: 10.1051/0004-6361/201015127

\bibitem[{{He{\l}miniak} {et~al.}(2024){He{\l}miniak}, {Olszewska},
  {Puciata-Mroczynska}, \& {Pawar}}]{hel24}
{He{\l}miniak}, K.~G., {Olszewska}, J.~M., {Puciata-Mroczynska}, M., \&
  {Pawar}, T.~B., {High-resolution spectroscopy of detached eclipsing binaries
  during total eclipses}. 2024, {\it \aap}, {\bf 691}, A170, DOI:
  10.1051/0004-6361/202450607

\bibitem[{{Moharana} {et~al.}(2024){Moharana}, {He{\l}miniak}, {Marcadon},
  {Pawar}, {Pawar}, {Konacki}, {Jord{\'a}n}, {Brahm}, \& {Espinoza}}]{moha24}
{Moharana}, A., {He{\l}miniak}, K.~G., {Marcadon}, F., {et~al.}, {Spectroscopy
  of eclipsing compact hierarchical triples: I. Low-mass double-lined and
  triple-lined systems}. 2024, {\it \aap}, {\bf 690}, A153, DOI:
  10.1051/0004-6361/202450797

\bibitem[{{Moharana} {et~al.}(2023){Moharana}, {He{\l}miniak}, {Marcadon},
  {Pawar}, {Konacki}, {Ukita}, {Kambe}, \& {Maehara}}]{moha23}
{Moharana}, A., {He{\l}miniak}, K.~G., {Marcadon}, F., {et~al.}, {Detached
  eclipsing binaries in compact hierarchical triples: triple-lined systems
  BD+442258 and KIC 06525196}. 2023, {\it \mnras}, {\bf 521}, 1908, DOI:
  10.1093/mnras/stad622

\bibitem[{{Pawar} {et~al.}(2024){Pawar}, {He{\l}miniak}, {Moharana}, {Pawar},
  {Pyatnytskyy}, {Lala}, \& {Konacki}}]{bpaw24}
{Pawar}, T.~B., {He{\l}miniak}, K.~G., {Moharana}, A., {et~al.}, {A
  comprehensive study of five candidate {\ensuremath{\delta}} Scuti-type
  pulsators in detached eclipsing binaries}. 2024, {\it \aap}, {\bf 691}, A101,
  DOI: 10.1051/0004-6361/202451126

\bibitem[{{Pojmanski}(2002)}]{asas}
{Pojmanski}, G., {The All Sky Automated Survey. Catalog of Variable Stars. I.
  0~h - 6~h Quarter of the Southern Hemisphere}. 2002, {\it \actaa}, {\bf 52},
  397, DOI: 10.48550/arXiv.astro-ph/0210283

\bibitem[{{Ratajczak} {et~al.}(2013){Ratajczak}, {He{\l}miniak}, {Konacki}, \&
  {Jord{\'a}n}}]{rata13}
{Ratajczak}, M., {He{\l}miniak}, K.~G., {Konacki}, M., \& {Jord{\'a}n}, A.,
  {Orbital and physical parameters of eclipsing binaries from the ASAS
  catalogue - V. Investigation of subgiants and giants: the case of ASAS
  J010538-8003.7, ASAS J182510-2435.5 and V1980 Sgr}. 2013, {\it \mnras}, {\bf
  433}, 2357, DOI: 10.1093/mnras/stt906

\bibitem[{{Ratajczak} {et~al.}(2016){Ratajczak}, {He{\l}miniak}, {Konacki},
  {Smith}, {Koz{\l}owski}, {Espinoza}, {Jord{\'a}n}, {Brahm}, {Hempel},
  {Anderson}, \& {Hellier}}]{rata16}
{Ratajczak}, M., {He{\l}miniak}, K.~G., {Konacki}, M., {et~al.}, {Orbital and
  physical parameters of eclipsing binaries from the ASAS catalogue - IX.
  Spotted pairs with red giants}. 2016, {\it \mnras}, {\bf 461}, 2234, DOI:
  10.1093/mnras/stw1448

\bibitem[{{Ratajczak} {et~al.}(2021){Ratajczak}, {Paw{\l}aszek},
  {He{\l}miniak}, {Konacki}, {Sybilski}, {Koz{\l}owski}, {Litwicki}, {Smith},
  {Miko{\l}ajczyk}, {Anderson}, \& {Hellier}}]{rata21}
{Ratajczak}, M., {Paw{\l}aszek}, R.~K., {He{\l}miniak}, K.~G., {et~al.},
  {Orbital and physical parameters of eclipsing binaries from the ASAS
  catalogue - XI. CHIRON investigation of long-period binaries}. 2021, {\it
  \mnras}, {\bf 500}, 4972, DOI: 10.1093/mnras/staa3488

\bibitem[{{Southworth}(2015)}]{sout15}
{Southworth}, J., {DEBCat: A Catalog of Detached Eclipsing Binary Stars}. 2015,
  in Astronomical Society of the Pacific Conference Series, Vol. {\bf  496},
  {\it Living Together: Planets, Host Stars and Binaries}, ed. S.~M.
  {Rucinski}, G.~{Torres}, \& M.~{Zejda}, 164

\end{thebibliography}

\end{document}